\documentclass[reprint,aip,jap,amsmath,amssymb,amsfonts]{revtex4-1}
\usepackage[utf8]{inputenc}
\usepackage{color}
\usepackage{multirow}
\usepackage[caption=false]{subfig}

\usepackage{graphicx}
\usepackage[dvipdfm,colorlinks=true,allcolors=blue]{hyperref}
\usepackage[capitalize]{cleveref}
\begin{document}
\author{Christopher B. McKitterick}
\affiliation{\mbox{Departments of Physics and Applied Physics, Yale University, New Haven, Connecticut 06520, USA}}
\author{Daniel E. Prober}
\affiliation{\mbox{Departments of Physics and Applied Physics, Yale University, New Haven, Connecticut 06520, USA}}
\author{Boris S. Karasik}
\affiliation{\mbox{Jet Propulsion Laboratory, California Institute of Technology, Pasadena, California 91109, USA}}
\title{Performance of Graphene Thermal Photon Detectors}
\begin{abstract}
We analyze the performance of graphene microstructures as thermal photon detectors and deduce the range of parameters that define a linear response. The saturation effects of a graphene thermal detector that operates beyond the linear range are described in detail for a single-photon detector (calorimeter). We compute the effect of operating beyond this linear range and find that sensitive detection occurs for such non-linear operation. We identify the optimum conditions and find that single-photon detection at terahertz (THz) frequencies should be feasible.
\end{abstract}
\maketitle
\section{Introduction}
Modern photon detectors are widely employed in sensitive applications ranging from astrophysical observations\cite{Benford2004,Karasik2005} to quantum communications.\cite{Hadfield2009} Some of the most sensitive detectors employ a thermal detection mechanism, where the photon energy (or power) is converted to a temperature increase. This temperature increase is sensed through a change of the electrical properties, such as a resistance change for a superconducting device biased on its superconducting transition. Transition-edge sensors (TESs)\cite{Cabrera2008} are well developed and their sensitivity can approach the fundamental limits for thermal detectors. \cite{Mather1984} Hot-electron TES detectors, in which only the electrons are heated, have achieved single-photon sensitivity in the near-infrared \cite{Lita2008} and the mid-IR.\cite{Karasik,Karasik2011a} Other detectors (such as quantum capacitance\cite{Stone2012} and kinetic inductance\cite{Day2003} detectors) are also being developed for these applications.

Lower-dimension carbon nanostructures (graphene and individual carbon nanotubes) are being considered as thermal photon detectors.\cite{Vora2011,Yan2011,Betz2012,Fong2012,Santavicca2011,Chudow2012,Fu2008} Since these systems have very few electrons that are active in the conduction process, the photon energy can cause significant heating. In addition, the very stiff carbon lattice should ensure that the electron-phonon interaction is weak, so cooling by phonon emission is small.


In this paper we investigate the limits on the sensitivity of graphene-based thermal detectors which employ a Johnson-noise temperature readout. Neither the dynamic range of detector operation nor the effect of operating beyond the linear range of thermal response have been studied previously. We concentrate on graphene, as carbon nanotubes have large quantum contact resistance and small shunting capacitance of the contact,\cite{Chudow2012} which makes THz photon coupling a challenge. Also, graphene has been shown to have simple Drude behavior of the conductivity from DC into the THz range,\cite{Horng2011} so modeling the absorption is feasible. The graphene detector exhibits saturation of the response when operated beyond the linear range, as defined below. We conclude that, with a Johnson-noise readout, operation beyond the linear range (near-equilibrium) provides the most sensitive detection and a graphene calorimeter can detect individual THz photons. Our study shows the limits and constraints of the Johnson-noise temperature readout. In this paper we first introduce the qualities of a desirable calorimeter and summarize the thermal properties of graphene. Then we describe how accurately one can read the temperature of graphene, both when the system is near equilibrium and after absorption of a 1-THz photon. Finally, we assess the use of graphene as a THz photon counter.
\section{Graphene as a Calorimeter}	

\begin{figure}
    \centering
    \begin{tabular}{cc}
    \multirow{-2}[15.75]{.5\columnwidth}{\subfloat[\label{fig:1a}]{\includegraphics[width=.5\columnwidth]{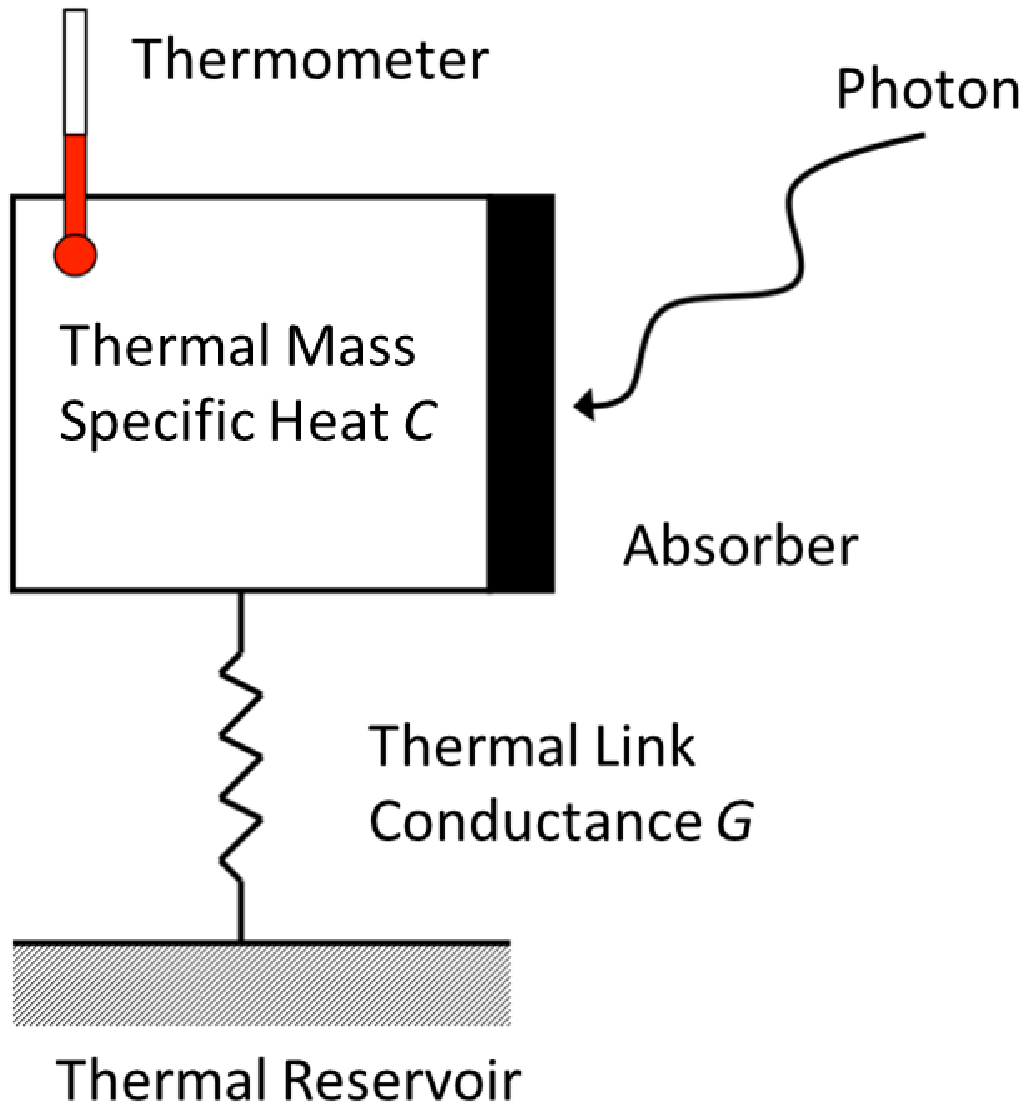}}}&
    \subfloat[\label{fig:1b}]{\includegraphics[width=.5\columnwidth]{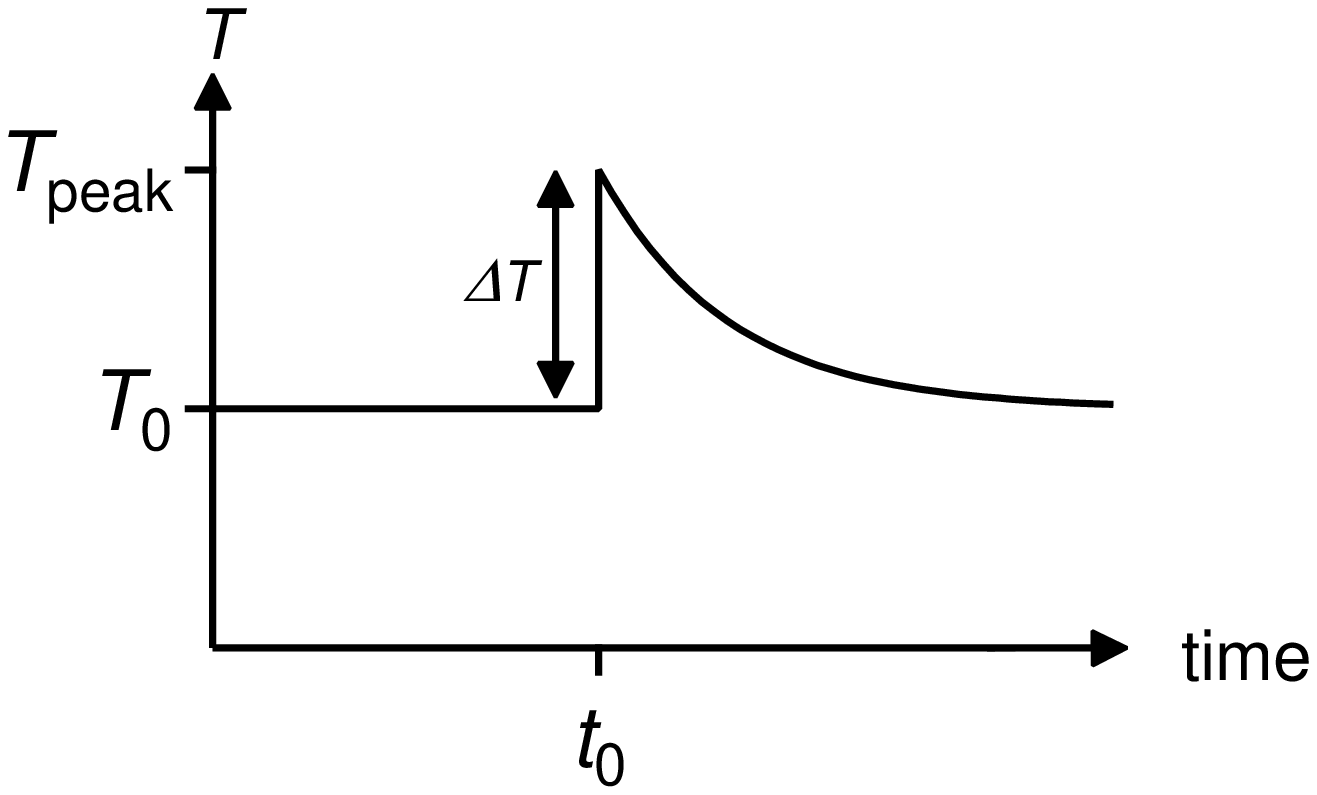}}\\
    &\subfloat[\label{fig:1c}]{\includegraphics[width=.35\columnwidth]{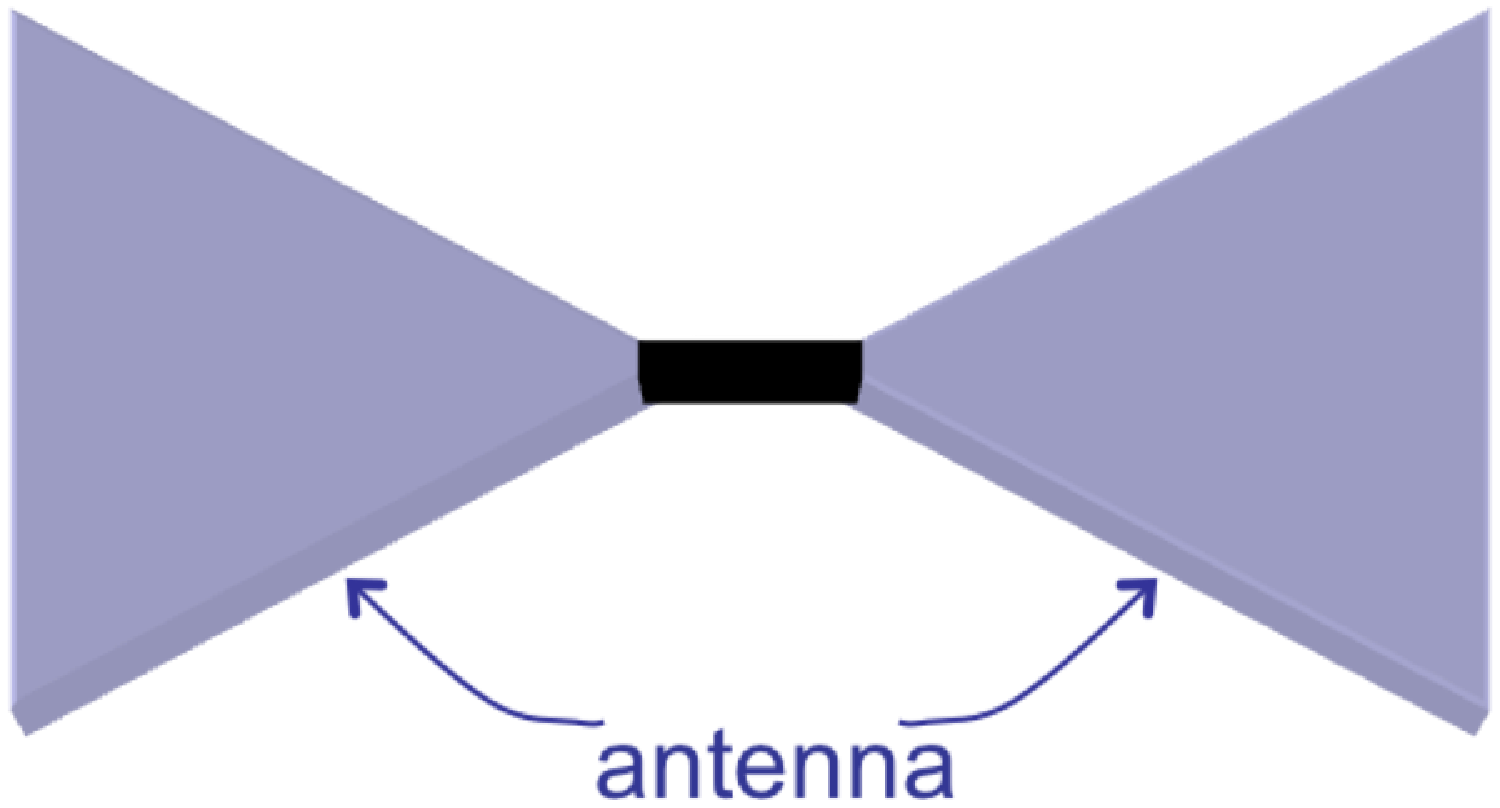}}
\end{tabular}
\caption{Schematic of bolometer or calorimeter. $C$ is the heat capacity and $G$ is the thermal conductance. In a graphene detector, $C$ and the thermometer function are provided by the electron subsystem. $G$ is provided by the coupling to the contacts and substrate as well as emitted Johnson noise. (b) Temperature response, $T(t)$, for a calorimeter. For graphene, resistance is nearly independent of temperature. (c) Schematic of antenna coupling to a small graphene device at the center. For a THz detector, the antenna linear dimension is a few 100 $\mu$m. The low frequency contacts are not shown.}\label{fig:1}
\end{figure}
A calorimeter works by absorbing a photon and as a result, changing its temperature, which is then read out to determine the photon energy. Such a thermal detector can also be used to measure the incident power due to a flux of photons, in which case it is called a bolometer. Whether it is optimal for power detection to operate in the integrating mode or in the photon counting mode is determined by the detector response time and the photon arrival rate, provided the detector can detect a single photon.\cite{Karasik2005} The figure of merit for a calorimeter is the energy resolution of the detector -- the ratio of photon energy to the fluctuations in energy during a measurement. In the linear range of response for a thermal detector, the intrinsic resolution is not affected by the size of the input signal. We define this linear range by requiring that the initial temperature increase due to a single absorbed photon (\cref{fig:1b}) be $\Delta T$ = ($T_{\mathrm{peak}}–T_0 ) \ll T_0$; $T_0$ is the device temperature with no photons and $T_\mathrm{peak}$ is the electron temperature shortly after photon absorption. For devices based on graphene, we require $\Delta T < \alpha T_0$ to specify the linear range, and in this work we set the limit $\alpha\approx$ 0.1 as a reasonable choice. 

Before discussing the performance of graphene as a photon detector, we first review the thermal properties of graphene. At low temperatures the heat capacity, $C$, is due to the electron specific heat and scales with temperature as $C=\gamma A T$, where $\gamma$ is the Sommerfeld coefficient of the electron system, and the thermal conductance scales as $G_\mathrm{eph}\propto A T^3$ for phonon cooling; $A$ is the graphene area.\cite{Voutilainen2011} These results are for finite electron densities away from the Dirac point in the standard model.\cite{Tse2009} Cooling by low-frequency photon emission can also contribute to the thermal conductance as $G_\mathrm{photon}\approx k_\mathrm{B}B$, where $B\ll k_B T/h$ is the coupled bandwidth to an impedance-matched load.\cite{Schmidt2004} In our case, we consider graphene which is impedance matched to an amplifier. Thus, there is no amplifier backaction (no electro-thermal feedback).\cite{Irwin1995} We assume that electron-electron interactions rapidly convert all the photon energy $E$ to electron thermal excitations at an initial temperature $T_\mathrm{peak}$ above $T_0$. Indeed, very fast electron-electron energy exchange for graphene has been reported.\cite{Voutilainen2011} 

If outdiffusion of heat through the contacts\cite{Prober1993} is suppressed, the thermal conductance $G$ is due to the emission of phonons and microwave photons\cite{Schmidt2004} by the heated electrons:
\begin{equation} G=G_\mathrm{eph}+G_\mathrm{photon}.\end{equation} 
Suppression of diffusion cooling out the contacts is desirable for a high sensitivity detector\cite{Vora2011,Fong2012} to allow a longer averaging time of the output signal. This can be achieved by transparent contact between graphene and superconducting leads,\cite{Borzenets2011,Du2008a,Ojeda-Aristizabal2009} where the superconductor energy gap prevents the outflow of thermal excitations from the graphene.\cite{Tinkham2004} Opaque (tunneling) superconducting contacts (superconductor-tunnel insulator-graphene) also prevent heat outdiffusion.\cite{Vora2011,Tinkham2004} With such contacts, the THz coupling can be efficient due to the finite tunnel junction capacitance. Additionally, the low-frequency resistance is temperature dependent, which provides a resistance readout of the graphene temperature change. We do not discuss that readout approach in this study. Instead, we treat graphene devices where the resistance (at the readout frequency) depends only weakly on temperature\cite{DasSarma2011} so the temperature is read out by measuring the Johnson noise. Because $dR/dT$ is very small, reading out the temperature change by measuring Johnson noise gives better sensitivity than readout using the resistance change with a finite bias current.\cite{Fong2012}
  
The expected theoretical form \cite{Tse2009} of $G_\mathrm{eph} = 4A\Sigma T^3$ is seen in the two experiments that measured electron-phonon coupling at low temperatures\cite{Betz2012,Fong2012} (above 2~K). $\Sigma$ measures the strength of electron-phonon coupling. Values of $\Sigma$ with moderate electron density $n\approx 2 \times 10^{11}$~cm$^{-2}$ are reported to be $\Sigma = 70$~mW/K$^4$m$^2$ for graphene on SiO$_2$\cite{Fong2012} and, with $n\approx 10^{12}$~cm$^{-2}$, $\Sigma=0.5 – 2$~mW/K$^4$m$^2$ for graphene on boron nitride (BN);\cite{Betz2012} the values of $\Sigma$ on BN are smaller by a factor of up to 140, so the choice of substrate material may affect $\Sigma$. $\Sigma$ scales with electron density as $\sqrt{n}$ for electron densities far from the charge neutrality point (as in the case of these measurements); this does not explain the difference in measured values of $\Sigma$. Effects of electron elastic scattering may significantly reduce the electron-phonon coupling in such graphene samples below 1~K.\cite{Chen2012} Since there are no measurements of $G_\mathrm{eph}$ below 2~K, we take the measured $T^3$ form to compute $G_\mathrm{eph}$. 
\subsection{Near-Equilibrium Device Noise}
We first consider the noise in the graphene device at or near equilibrium, where $T\approx T_0$. There are two dominant sources of noise. The first is due to intrinsic energy fluctuations in the device. These fluctuations can be thought of as energy carriers (i.e., photons, phonons) entering and exiting the graphene device at random.\cite{Mather1982} The second source of noise is the accuracy limit with which one can measure the temperature of a device using Johnson-noise thermometry.\cite{Dicke1946}

\subsubsection{Intrinsic Noise}
The rms intrinsic energy fluctuation of the detector, $\delta E_\mathrm{intr}$, is due to the intrinsic thermodynamic fluctuations. For measurements using the thermal response bandwidth corresponding to the thermal time constant, $\tau=C/G$, this is given by\cite{Mather1982}
\begin{align}
\delta E_\mathrm{intr}&=(k_\mathrm{B}T^2C)^{1/2}\\
\delta T_\mathrm{intr}&=(k_\mathrm{B}T^2/C)^{1/2}.
\label{equ:rmsintTE}
\end{align}
Here $C$ is the heat capacity at temperature $T$; at $T_0$, $C=C_0$ and $G=G_0$ . $\delta T_\mathrm{intr}$ is the rms equivalent temperature fluctuation. We find in the following section that readout noise is larger than this intrinsic noise. For that case, the optimum averaging time is indeed $\tau_\mathrm{avg}=C/G$.


The intrinsic energy resolving power (often called the energy resolution) $R_\mathrm{intr}$ is usually defined using the energy fluctuation (full width at half maximum, fwhm), which is equal to $2\sqrt{2\ln 2}\cdot\delta E_\mathrm{intr}\approx 2.35\cdot\delta E_\mathrm{intr}$, where $\delta E_\mathrm{intr}$ is the rms fluctuation. Thus, at $T=T_0$, 
\begin{equation}
R_\mathrm{intr}(\mathrm{fwhm})\approx 0.42\cdot E/\sqrt{k_\mathrm{B}T_0^2C_0}.\
\label{equ:Rintfwhm}
\end{equation}
However, for the measurement scheme considered, we read out the temperature of the device, not the energy. While $\delta T$ and $\delta E$ give equivalent information in the linear range, we find that sensitive detection of THz photons occurs far beyond near-equilibrium range. As a result, the resolution in \cref{equ:Rintfwhm} is the relevant figure of merit only for operation in the linear range. A criterion based on temperature measurement will be developed in \cref{sec:non-equil} that is relevant for operation beyond the linear range.

\subsubsection{Amplifier Noise}
We now address the issue of amplifier noise and focus specifically on readout of the temperature using the emitted Johnson noise. This readout measures the emitted noise power over a time $\tau_\mathrm{avg}$. The increase in the average emitted noise power at frequencies $f < k_\mathrm{B}T/h$ is given by $P(t)= k_\mathrm{B}\Delta T(t)B$, where $B$ is the observing bandwidth. We assume that the noise emission is characterized simply with an instantaneous electron temperature,\cite{Voutilainen2011,Burke1996,Dicke1946} that the detector resistance is impedance matched to the amplifier (achievable with $n\approx 10^{12}~$cm$^{-2}$ and a wide graphene flake), and that we can use the low-frequency limit of the Johnson noise.\cite{Callen1951} Thus, the signal measured in the absence of noise is given by
\begin{align}
\nonumber S &= \int_0^{\tau_\mathrm{avg}} k_\mathrm{B} \Delta T(t)Bdt,
\end{align}
where $B$ is the output bandwidth and $\Delta T(t)=T(t)-T_0$. If we take $\tau_\mathrm{avg}$ to be equal to the thermal time constant, $\tau$, this becomes:
\begin{equation}
S= k_\mathrm{B}\Delta T_\mathrm{avg}B\tau,
\label{equ:sigT}
\end{equation}
where $\Delta T_\mathrm{avg}=\frac{1}{\tau}\int_0^\tau (T(t)-T_0)dt.$

The rms accuracy with which the electron temperature $T$ can be measured via Johnson noise in a time $\tau$ is\cite{Dicke1946}
\begin{subequations}
\begin{equation} 
\delta T_\mathrm{readout} = \frac{T_\mathrm{a} + T}{\sqrt{B\tau}},
\label{equ:Treadout}
\end{equation}
where $T_\mathrm{a}$ is the noise temperature of the amplifier.\cite{Letzter1970}
Near equilibrium, $T\approx T_0$ and \cref{equ:Treadout} is equivalent to an rms energy fluctuation of
\begin{equation}
\delta E_\mathrm{readout} = C_0\frac{T_\mathrm{a} + T_0}{\sqrt{B\tau}}.
\end{equation}
\end{subequations}

The optimum values for $B$ and $T_\mathrm{a}$ require careful consideration, and are constrained by amplifier choice. For the Johnson-noise readout, large bandwidth is desirable if phonon cooling, $G_\mathrm{eph}$, dominates. This reduces $\delta T_\mathrm{readout}$. However, because $G_\mathrm{photon}$ is proportional to $B$ and $\tau=C/G$, once $G_\mathrm{photon}>G_\mathrm{eph}$, further increasing the bandwidth has limited benefit. To provide specific examples of amplifiers, we will consider two amplifiers for the remaining calculations. The first (referred to as amplifier A) is a hypothetical near-quantum-limited parametric amplifier (paramp) which has a center frequency $f_0$, bandwidth $B$, and noise temperature $T_\mathrm{a}$ that have all been scaled down by a factor of approximately 10 from published data.\cite{HoEom2012}
Although the paramp specifications listed in \cref{table:amp} have not been demonstrated, the measured results at frequencies approximately 10 times larger provide encouragement for their possible realization.
The second (amplifier B) is a SQUID (Superconducting Quantum Interference Device) amplifier with its measured performance.~\cite{DeFeo2012} The amplifier parameters are summarized in \cref{table:amp}.
\begin{table}
\centering
\resizebox{\columnwidth}{!}{
\begin{tabular}{llll}
\toprule
Amplifier\hspace{5mm} & Noise Temperature, $T_\mathrm{a}$\hspace{5mm} & Bandwidth, $B$\hspace{5mm} & Center Frequency, $f_0$\\\colrule
A\cite{HoEom2012}&\multirow{2}{*}{\large{0.15~K}}&\multirow{2}{*}{\large{150~MHz}}&\multirow{2}{*}{\large{1~GHz}}\\
paramp&&&\\\colrule
B\cite{DeFeo2012}&\multirow{2}{*}{\large{0.6~K}}&\multirow{2}{*}{\large{150~MHz}}&\multirow{2}{*}{\large{3.9~GHz}}\\
SQUID&&&\\\botrule
\end{tabular}
}
\caption{Summary of amplifier characteristics for the parametric and SQUID amplifiers.}
\label{table:amp}
\end{table}

The total resolving power is limited by both intrinsic thermodynamic fluctuations and the accuracy limits of the temperature measurement:
\begin{align}
\nonumber \delta T_\mathrm{tot}&=\sqrt{\delta T_\mathrm{intr}^2+\delta T_\mathrm{readout}^2}\\
\nonumber \delta E_\mathrm{tot}&=C_0\cdot\delta T_\mathrm{tot}
\end{align}
\begin{figure}[h!]
 \centering
 \includegraphics[width=.8\columnwidth]{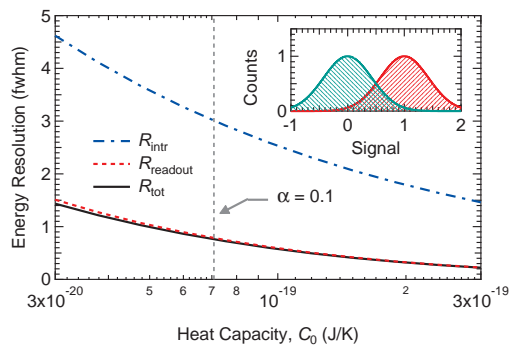}
 \caption{Prediction of energy resolution for a graphene detector operating in the linear range with $\Sigma=0.5~$mW/K$^4$m$^2$; for a 1-THz photon, this is for $C_0\geq 7\times 10^{-20}$~J/K, which corresponds to $\alpha\leq 0.1$ for $T_0=0.1$~K. The readout is performed using amplifier A. The inset shows a
normalized histogram of recorded signals from single-photon detection events and from sampling the baseline, with $R_\mathrm{tot} = 1$. This corresponds to $\alpha=0.14$.}
\label{fig:linearres-QL}
\end{figure}
Taking a large enough heat capacity to remain in the linear range, which we define as $\Delta T<0.1 T_0$, we find that the resulting calculated fwhm energy resolution, 
\begin{equation}R_\mathrm{tot}~(\mathrm{fwhm})=0.42\cdot E/\delta E_\mathrm{tot},
\end{equation}
is less than 1 with amplifier A. In \cref{fig:linearres-QL}, we plot this energy resolution in the linear range and also present the normalized histograms that would result from counting 1-THz photons with $R_\mathrm{tot}=1$ and from sampling the baseline with no photon events.  From this figure we see that it is not possible to clearly distinguish between the ensemble of zero photons, centered on 0 signal, and the one photon histogram centered on 1. For the realistic case where the number of zero-photon events is much greater than the number of one-photon events, the zero-photon peak would be much larger and the overlap of the two histograms would be worse. We therefore conclude that we must consider non-linear operation. Such operation increases $\Delta T$ more than it increases $\delta T_\mathrm{readout}$, which remains the dominant source of broadening.


\subsection{Non-Equilibrium Device Noise}
\label{sec:non-equil}
In the previous section, we considered  the device noise with no incident photons. However, there do not appear to be any set of device parameters for $T\geq 0.1~$K that provide good sensitivity and keep operation in the linear range.
We thus need to consider operation beyond the linear range, for which the arrival of a photon will significantly heat the graphene. The temperature rise is computed from the electron energy, given by $U = (\gamma/2) AT^2$ at low temperature,\cite{Kittel2004} where $T$ is the electron temperature. Thus, for photon absorption, $\Delta U = E = (\gamma/2)A(T_\mathrm{peak}^2–T_0^2)$.


\begin{figure}[h!]
 \centering
 \includegraphics[width=.8\columnwidth]{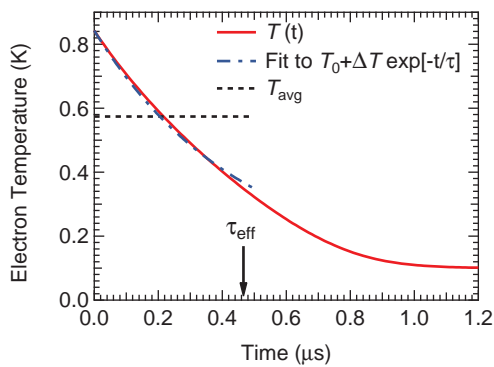}
 \caption{Results from calculation of $T(t)$ at $C_0=2\times 10^{-22}$~J/K and $B=150$ MHz with $\Sigma=0.5~$mW/K$^4$m$^2$ after absorbing a 1-THz photon. Also shown is the fit of the temperature to an exponential function, with $T_0=0.1~$K and $\Delta T(t)$ calculated as discussed in the text. The effective time constant extracted is $\tau_\mathrm{eff}=460~$ns.}
\label{fig:tempvtime}
\end{figure}

\begin{figure}[h!]
 \centering
 \includegraphics[width=.8\columnwidth]{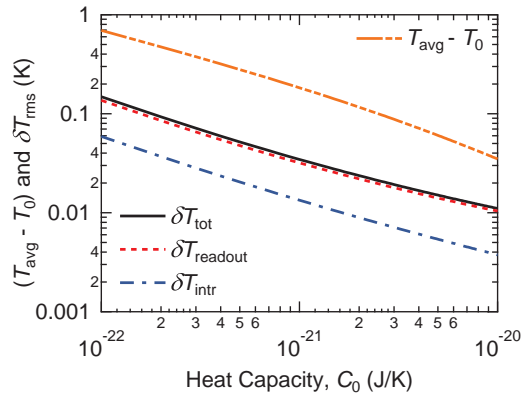}
 \caption{Prediction of rms temperature fluctuations when excited by a 1-THz photon for $T_0=0.1$~K and $\Sigma=0.5~$mW/K$^4$m$^2$. The readout calculation uses amplifier A, described in the text. $T_\mathrm{avg}$ is the average temperature of the graphene over $\tau_\mathrm{eff}$ following the arrival of a 1-THz photon. For $C_0=2\times 10^{-22}$~J/K, $T_\mathrm{avg}-T_0=0.48~$K.}
\label{fig:nonequilres-QL}
\end{figure}

To determine $\delta E_\mathrm{intr}$ outside the linear range the detailed time evolution of the temperature $T(t)$ is considered. In the linear range $T(t)$ is given by $\Delta T(t) = T(t) - T_0 = (E/C_0) \exp(-\tau_0/t)$ for $t\geq 0$; $\tau_0 = C_0/G_0$.  
For large $\Delta T$, the time decay is not a simple exponential function. The initial time decay at $T_\mathrm{peak}$ is fast if phonon emission is dominant ($\tau\propto T^{-2}$ for phonon cooling), but the later decay of $\Delta T(t)$ back to $T_0$ occurs more slowly. Additionally, for large output bandwidths, much of the cooling is done through emitted Johnson noise. In order to determine the electron temperature as a function of time, $T(t)$, (shown in \cref{fig:tempvtime}) we solve numerically 
\begin{align}
\nonumber \frac{dU}{dt}&=\gamma AT\frac{dT}{dt}\\
\nonumber &=-P(T)\\
\nonumber &=-\Sigma A(T^4-T_0^4)-k_\mathrm{B}B(T-T_0)\\
\Rightarrow -dt&=\frac{\gamma AT dT}{\Sigma A(T^4-T_0^4)+k_\mathrm{B}B(T-T_0)}.
\label{eq:diffte}
\end{align}
The computed $T(t)-T_0$ is then fit to an exponential decay to determine an approximate effective time constant, $\tau_\mathrm{eff}$. $T_\mathrm{avg}$ is determined from this solution as shown in \cref{fig:tempvtime} using the weaker electron-phonon coupling, $\Sigma=0.5~$mW/K$^4$m$^2$. Using \cref{equ:rmsintTE} we can directly calculate the intrinsic thermal fluctuations. The temperature $T$ used in \cref{equ:rmsintTE} is given with reasonable accuracy by $(T_\mathrm{avg}+T_0)/2$, as the temperature fluctuations are due to phonons or photons leaving the graphene (at $T_\mathrm{avg}$) or entering from the substrate (at $T_0$). The heat capacity of the graphene is given by $C(T_\mathrm{avg})$. We use the average electron temperature $T_\mathrm{avg}$ to specify $T$ in \cref{equ:Treadout}. The prediction of the temperature fluctuations for this average temperature is shown in \cref{fig:nonequilres-QL} as a function of $C_0$, for $T_0=0.1~K$ with $\Sigma=0.5~$mW/K$^4$m$^2$.
In \cref{fig:nonequilres-QL} we see that $\delta T_\mathrm{readout}>\delta T_\mathrm{intr}$. Thus, even though our evaluation of $\delta T_\mathrm{intr}$ may not be exact, the readout noise dominates and determines $\delta T_\mathrm{tot}$, as seen in \cref{fig:nonequilres-QL}. The rms fluctuation widths in \cref{fig:nonequilres-QL} apply for absorption of individual 1-THz photons. 

\begin{figure}[h!]
 \centering
 \includegraphics[width=.8\columnwidth]{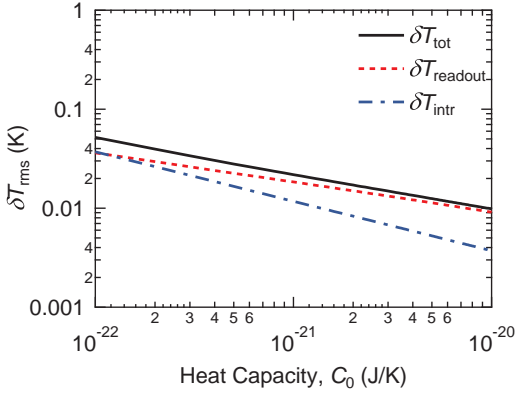}
 \caption{Prediction of rms near-equilibrium temperature fluctuations with no photons for $T_0=0.1$~K and $\Sigma=0.5~$mW/K$^4$m$^2$. The readout is performed using amplifier A.}
\label{fig:equilres-QL}
\end{figure}

In \cref{fig:equilres-QL}, we plot the equivalent temperature fluctuations for the near-equilibrium case; this is relevant for the photon detector when there are no absorbed photons. In both \cref{fig:nonequilres-QL,fig:equilres-QL}, $\tau_\mathrm{eff}$ depends on $C_0$. The $\tau_\mathrm{eff}$ used to calculate each value of $\delta T_\mathrm{readout}$ and $T_\mathrm{avg}$ is different for each heat capacity and is determined from fits analogous to the one shown in \cref{fig:tempvtime}. The same values are used in \cref{fig:nonequilres-QL,fig:equilres-QL}.

\begin{figure}[h!]
 \centering
 \includegraphics[width=.9\columnwidth]{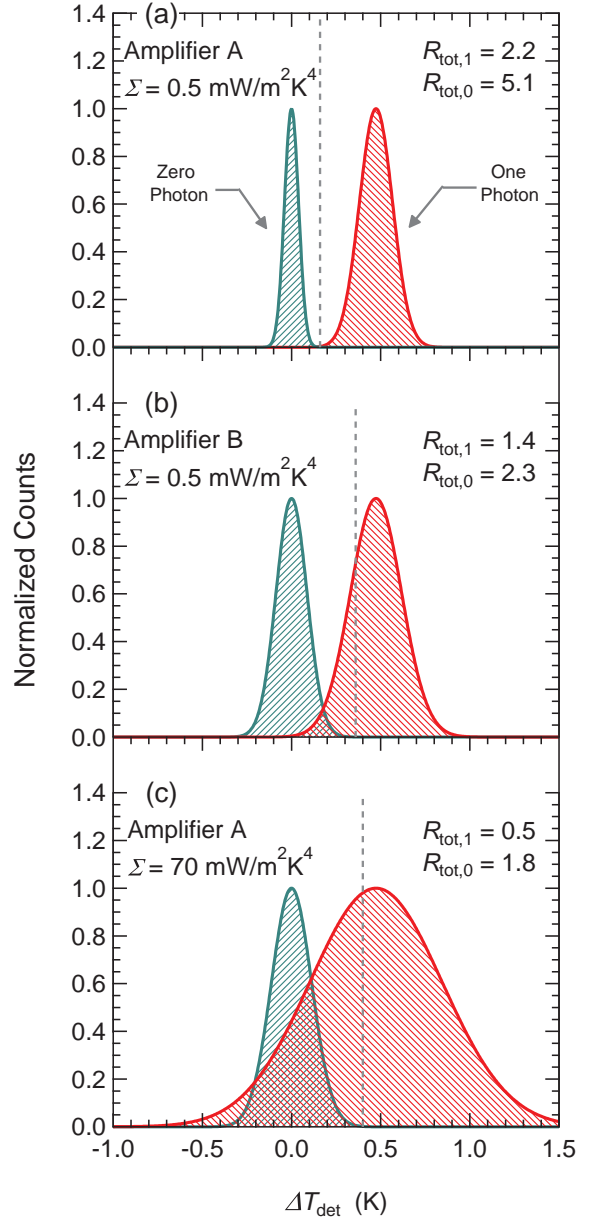}
 \caption{Normalized histograms with $C_0=2\times 10^{-22}~$J/K for photon counts using (a) amplifier A and (b) amplifier B. (c) shows the predicted photon count distribution if the larger value for $\Sigma$ is assumed, with amplifier A. The dashed line in each figure indicates the estimated threshold temperature, $\Delta T_\mathrm{det,min}$, that is necessary to meet the single-photon detection requirements described in \cref{sec:application}.}
\label{fig:histograms}
\end{figure}

The energy resolving power will be evident if one plots a histogram of photon absorption events for a large ensemble of events. The histogram records the detected temperature rise for each absorbed photon. We denote this detected temperature rise as $\Delta T_\mathrm{det}=S/(k_BB\tau_\mathrm{avg}),$ with the measured value of $S$ for each count. The ensemble average of the $T_\mathrm{det}$ values is $T_\mathrm{avg}$ and the fwhm width of the distribution is $2.35\cdot \delta T_\mathrm{tot}$ since $\delta T_\mathrm{tot}$ as defined above is an rms value. We plot this normalized distribution of single-photon counts in \cref{fig:histograms} for $\Delta T_\mathrm{det}=T_\mathrm{det}-T_0$. The functional form is 
\begin{equation}
\mathrm{Counts}\propto \exp\left[-\frac{(T_\mathrm{det}-T_\mathrm{avg})^2}{2\cdot\delta T_\mathrm{tot}^2}\right].
\end{equation}
For these plots, we choose $C_0=2\times 10^{-22}~$J/K. This gives a larger $\Delta T_\mathrm{avg}$ than is obtained for larger values of $C_0$. We do not consider a smaller heat capacity as it would lead to more significant heating of the electron system, potentially allowing the high energy tail of the hot electron distribution to diffuse over the energy gap of the superconducting contacts. With a carrier density of $10^{12}~$cm$^{-2}$ this value of $C_0=2\times 10^{-22}~$J/K corresponds to an area approximately equal to $4.5~\mu$m$^2$. Somewhat larger values of $C_0$ (and $A$) would have similar performance.

We also plot in \cref{fig:histograms} the predicted distribution if we sample the detector output with no input photons, using the same amplifier and $\tau_\mathrm{eff}$ as employed for photon detection. These are the ``zero-photon'' events, centered around $\Delta T_\mathrm{det}=0$. In the figure, the curves are normalized to the same height. However, in practice there will be many more zero-photon than one-photon events in the applications for which these devices are being considered.\cite{Karasik2005,Day2003} Assuming one samples at a rate equal to $\tau_\mathrm{avg}^{-1}$, a signal photon arrival rate\cite{Karasik2005} of $N=10^4$~s$^{-1}$ and $\tau_\mathrm{avg}\sim 0.5~\mu$s, we expect there to be approximately 200 more zero-photon events than one-photon events. For these plots, we define the temperature resolution as $R_{\mathrm{tot},m}~\mathrm{(fwhm)}=\Delta T_\mathrm{avg}/(2.35\cdot \delta T_\mathrm{tot})$, where $m$ can be 0 or 1 to denote the zero-photon or one-photon case; $\delta T_\mathrm{tot}$ is the rms value.

\Cref{fig:histograms} clearly demonstrates that resolution is reduced upon the arrival of an incident photon. However, even the one-photon resolution is greater than is shown in the inset of \cref{fig:linearres-QL}. In that figure, the zero-photon and the one-photon histograms have the same width. With a small $C_0$, we find that the resolution is improved for single-photon detection and is significantly improved for the zero-photon histogram.

\section{Application of Graphene-Based Detectors}
\label{sec:application}
The previous section makes clear that the temperature resolution of a graphene calorimeter is not signal independent. After a photon has been absorbed, the fluctuations are significantly larger. The increase in fluctuation likely precludes the use of a graphene calorimeter for high-resolution THz spectroscopy or other methods of photon detection which require fine energy resolution. However, for single-photon counting, the outlook is positive.

One can use graphene as a single-photon ``click" detector in which a threshold temperature rise $\Delta T_\mathrm{det,min}$ is set. This threshold would be set above the main distribution of the equilibrium (zero-photon) fluctuations. We set $\Delta T_\mathrm{det,min}$ to exclude almost all events with a smaller signal. This reduces the dark count rate below the specified maximum, but may also exclude some actual (single-photon) events. The threshold needs to take into account the much larger number of zero-photon events than is displayed in the normalized histograms of \cref{fig:histograms}. 

To obtain a negligible dark count rate with the parametric amplifier in \cref{fig:histograms}a (amplifier A), one would need to set a threshold value for $\Delta T_\mathrm{det,min}$ of 0.16~K. This threshold is indicated by the dashed line in \cref{fig:histograms}a and it demonstrates that, with this choice of amplifier, nearly all single-photon events will be counted and almost all of the zero-photon events will be excluded. With the SQUID amplifier in \cref{fig:histograms}b, a similar restriction on dark counts would require a higher threshold ($\Delta T_\mathrm{det,min}=0.36~$K) and some single-photon events would be excluded. Worse performance is found by assuming the stronger electron-phonon coupling (shown in \cref{fig:histograms}c). A threshold $\Delta T_\mathrm{det,min}= 0.4~$K would need to be set to prevent excessive dark counts, excluding nearly half of the single-photon events. 

We conclude that a weaker value for electron-phonon coupling is necessary for an efficient detector using the Johnson noise readout. However, if the resistance change due to photon absorption is large, as would be the case with superconducting tunnel contacts, the predicted performance can be better. In this case the intrinsic energy fluctuations could dominate and one might approach the intrinsic energy resolution (fwhm) shown in \cref{fig:linearres-QL}.

\section{Conclusions}
Graphene detectors that use a Johnson-noise readout of detector temperature can count single THz photons with reasonable energy resolution. The design must carefully take account of the effects of photon heating on the performance, and should employ graphene with weak electron-phonon coupling. It will be necessary to use temperatures below 1~K and low-noise amplifiers with nearly quantum-limited sensitivity. These requirements are largely necessitated by the Johnson-noise readout and the fast thermal response time of graphene. With a temperature-independent resistance (such as that of graphene), measurement of the electron temperature is difficult. Thus, the implementation of such a detector will require real care.

Recent experiments\cite{Betz2012,Fong2012} have shown that for some graphene samples the electron-phonon thermal conductance can differ by up to two orders of magnitude. Mechanisms that can decrease the electron-phonon emission rate in graphene are also under study.\cite{Chen2012} Reduced phonon emission would increase the sensitivity for both single-photon and power detection. Thus, direct experimental tests and a more thorough understanding of how electrons interact with the lattice in real graphene samples will directly benefit the development of ultrasensitive photon detectors, and is one of the important current research challenges.

\begin{acknowledgments}
We acknowledge support from NSF DMR 0907082 and Yale University; we also acknowledge productive discussions with  F.W. Carter,  J.D. Chudow, H.D. Drew, X. Du, K.C. Fong, M. Fuhrer, S.H. Moseley D.F. Santavicca, A.V. Sergeev, and A.E. Szymkowiak. The research of B.S. Karasik was carried out at the Jet Propulsion Laboratory, California Institute of Technology, under a contract with the National Aeronautics and Space Administration.
\end{acknowledgments}


\begin{thebibliography}{37}%
\makeatletter
\providecommand \@ifxundefined [1]{%
 \@ifx{#1\undefined}
}%
\providecommand \@ifnum [1]{%
 \ifnum #1\expandafter \@firstoftwo
 \else \expandafter \@secondoftwo
 \fi
}%
\providecommand \@ifx [1]{%
 \ifx #1\expandafter \@firstoftwo
 \else \expandafter \@secondoftwo
 \fi
}%
\providecommand \natexlab [1]{#1}%
\providecommand \enquote  [1]{``#1''}%
\providecommand \bibnamefont  [1]{#1}%
\providecommand \bibfnamefont [1]{#1}%
\providecommand \citenamefont [1]{#1}%
\providecommand \href@noop [0]{\@secondoftwo}%
\providecommand \href [0]{\begingroup \@sanitize@url \@href}%
\providecommand \@href[1]{\@@startlink{#1}\@@href}%
\providecommand \@@href[1]{\endgroup#1\@@endlink}%
\providecommand \@sanitize@url [0]{\catcode `\\12\catcode `\$12\catcode
  `\&12\catcode `\#12\catcode `\^12\catcode `\_12\catcode `\%12\relax}%
\providecommand \@@startlink[1]{}%
\providecommand \@@endlink[0]{}%
\providecommand \url  [0]{\begingroup\@sanitize@url \@url }%
\providecommand \@url [1]{\endgroup\@href {#1}{\urlprefix }}%
\providecommand \urlprefix  [0]{URL }%
\providecommand \Eprint [0]{\href }%
\providecommand \doibase [0]{http://dx.doi.org/}%
\providecommand \selectlanguage [0]{\@gobble}%
\providecommand \bibinfo  [0]{\@secondoftwo}%
\providecommand \bibfield  [0]{\@secondoftwo}%
\providecommand \translation [1]{[#1]}%
\providecommand \BibitemOpen [0]{}%
\providecommand \bibitemStop [0]{}%
\providecommand \bibitemNoStop [0]{.\EOS\space}%
\providecommand \EOS [0]{\spacefactor3000\relax}%
\providecommand \BibitemShut  [1]{\csname bibitem#1\endcsname}%
\let\auto@bib@innerbib\@empty
\bibitem [{\citenamefont {Benford}\ and\ \citenamefont
  {Moseley}(2004)}]{Benford2004}%
  \BibitemOpen
  \bibfield  {author} {\bibinfo {author} {\bibfnamefont {D.~J.}\ \bibnamefont
  {Benford}}\ and\ \bibinfo {author} {\bibfnamefont {S.}~\bibnamefont
  {Moseley}},\ }\href {\doibase 10.1016/j.nima.2003.11.295} {\bibfield
  {journal} {\bibinfo  {journal} {Nuclear Instruments and Methods in Physics
  Research Section A: Accelerators, Spectrometers, Detectors and Associated
  Equipment}\ }\textbf {\bibinfo {volume} {520}},\ \bibinfo {pages} {379}
  (\bibinfo {year} {2004})}\BibitemShut {NoStop}%
\bibitem [{\citenamefont {Karasik}\ and\ \citenamefont
  {Sergeev}(2005)}]{Karasik2005}%
  \BibitemOpen
  \bibfield  {author} {\bibinfo {author} {\bibfnamefont {B.}~\bibnamefont
  {Karasik}}\ and\ \bibinfo {author} {\bibfnamefont {A.}~\bibnamefont
  {Sergeev}},\ }\href {\doibase 10.1109/TASC.2005.849963} {\bibfield  {journal}
  {\bibinfo  {journal} {IEEE Transactions on Appiled Superconductivity}\
  }\textbf {\bibinfo {volume} {15}},\ \bibinfo {pages} {618} (\bibinfo {year}
  {2005})}\BibitemShut {NoStop}%
\bibitem [{\citenamefont {Hadfield}(2008)}]{Hadfield2009}%
  \BibitemOpen
  \bibfield  {author} {\bibinfo {author} {\bibfnamefont {R.~H.}\ \bibnamefont
  {Hadfield}},\ }\href {\doibase 10.1038/nphoton.2009.230} {\bibfield
  {journal} {\bibinfo  {journal} {Nature Photonics}\ }\textbf {\bibinfo
  {volume} {3}},\ \bibinfo {pages} {696} (\bibinfo {year} {2008})}\BibitemShut
  {NoStop}%
\bibitem [{\citenamefont {Cabrera}(2008)}]{Cabrera2008}%
  \BibitemOpen
  \bibfield  {author} {\bibinfo {author} {\bibfnamefont {B.}~\bibnamefont
  {Cabrera}},\ }\href {\doibase 10.1007/s10909-007-9632-2} {\bibfield
  {journal} {\bibinfo  {journal} {Journal of Low Temperature Physics}\ }\textbf
  {\bibinfo {volume} {151}},\ \bibinfo {pages} {82} (\bibinfo {year}
  {2008})}\BibitemShut {NoStop}%
\bibitem [{\citenamefont {Mather}(1984)}]{Mather1984}%
  \BibitemOpen
  \bibfield  {author} {\bibinfo {author} {\bibfnamefont {J.~C.}\ \bibnamefont
  {Mather}},\ }\href {\doibase 10.1364/AO.23.000584} {\bibfield  {journal}
  {\bibinfo  {journal} {Applied Optics}\ }\textbf {\bibinfo {volume} {23}},\
  \bibinfo {pages} {584} (\bibinfo {year} {1984})}\BibitemShut {NoStop}%
\bibitem [{\citenamefont {Lita}, \citenamefont {Miller},\ and\ \citenamefont
  {Nam}(2008)}]{Lita2008}%
  \BibitemOpen
  \bibfield  {author} {\bibinfo {author} {\bibfnamefont {A.~E.}\ \bibnamefont
  {Lita}}, \bibinfo {author} {\bibfnamefont {A.~J.}\ \bibnamefont {Miller}}, \
  and\ \bibinfo {author} {\bibfnamefont {S.~W.}\ \bibnamefont {Nam}},\ }\href
  {\doibase 10.1364/OE.16.003032} {\bibfield  {journal} {\bibinfo  {journal}
  {Optics Express}\ }\textbf {\bibinfo {volume} {16}},\ \bibinfo {pages} {3032}
  (\bibinfo {year} {2008})}\BibitemShut {NoStop}%
\bibitem [{\citenamefont {Karasik}\ \emph {et~al.}(2012)\citenamefont
  {Karasik}, \citenamefont {Pereverzev}, \citenamefont {Soibel}, \citenamefont
  {Santavicca}, \citenamefont {Prober}, \citenamefont {Olaya},\ and\
  \citenamefont {Gershenson}}]{Karasik}%
  \BibitemOpen
  \bibfield  {author} {\bibinfo {author} {\bibfnamefont {B.~S.}\ \bibnamefont
  {Karasik}}, \bibinfo {author} {\bibfnamefont {S.~V.}\ \bibnamefont
  {Pereverzev}}, \bibinfo {author} {\bibfnamefont {A.}~\bibnamefont {Soibel}},
  \bibinfo {author} {\bibfnamefont {D.~F.}\ \bibnamefont {Santavicca}},
  \bibinfo {author} {\bibfnamefont {D.~E.}\ \bibnamefont {Prober}}, \bibinfo
  {author} {\bibfnamefont {D.}~\bibnamefont {Olaya}}, \ and\ \bibinfo {author}
  {\bibfnamefont {M.~E.}\ \bibnamefont {Gershenson}},\ }\href {\doibase
  10.1063/1.4739839} {\bibfield  {journal} {\bibinfo  {journal} {Applied
  Physics Letters}\ }\textbf {\bibinfo {volume} {101}},\ \bibinfo {pages}
  {052601} (\bibinfo {year} {2012})}\BibitemShut {NoStop}%
\bibitem [{\citenamefont {Karasik}, \citenamefont {Sergeev},\ and\
  \citenamefont {Prober}(2011)}]{Karasik2011a}%
  \BibitemOpen
  \bibfield  {author} {\bibinfo {author} {\bibfnamefont {B.~S.}\ \bibnamefont
  {Karasik}}, \bibinfo {author} {\bibfnamefont {A.~V.}\ \bibnamefont
  {Sergeev}}, \ and\ \bibinfo {author} {\bibfnamefont {D.~E.}\ \bibnamefont
  {Prober}},\ }\href {\doibase 10.1109/TTHZ.2011.2159560} {\bibfield  {journal}
  {\bibinfo  {journal} {IEEE Transactions on Terahertz Science and Technology}\
  }\textbf {\bibinfo {volume} {1}},\ \bibinfo {pages} {97} (\bibinfo {year}
  {2011})}\BibitemShut {NoStop}%
\bibitem [{\citenamefont {Stone}\ \emph {et~al.}(2012)\citenamefont {Stone},
  \citenamefont {Megerian}, \citenamefont {Day}, \citenamefont {Echternach},
  \citenamefont {Bueno},\ and\ \citenamefont {Llombart}}]{Stone2012}%
  \BibitemOpen
  \bibfield  {author} {\bibinfo {author} {\bibfnamefont {K.~J.}\ \bibnamefont
  {Stone}}, \bibinfo {author} {\bibfnamefont {K.~G.}\ \bibnamefont {Megerian}},
  \bibinfo {author} {\bibfnamefont {P.~K.}\ \bibnamefont {Day}}, \bibinfo
  {author} {\bibfnamefont {P.~M.}\ \bibnamefont {Echternach}}, \bibinfo
  {author} {\bibfnamefont {J.}~\bibnamefont {Bueno}}, \ and\ \bibinfo {author}
  {\bibfnamefont {N.}~\bibnamefont {Llombart}},\ }\href {\doibase
  10.1063/1.4731880} {\bibfield  {journal} {\bibinfo  {journal} {Applied
  Physics Letters}\ }\textbf {\bibinfo {volume} {100}},\ \bibinfo {pages}
  {263509} (\bibinfo {year} {2012})}\BibitemShut {NoStop}%
\bibitem [{\citenamefont {Day}\ \emph {et~al.}(2003)\citenamefont {Day},
  \citenamefont {LeDuc}, \citenamefont {Mazin}, \citenamefont {Vayonakis},\
  and\ \citenamefont {Zmuidzinas}}]{Day2003}%
  \BibitemOpen
  \bibfield  {author} {\bibinfo {author} {\bibfnamefont {P.~K.}\ \bibnamefont
  {Day}}, \bibinfo {author} {\bibfnamefont {H.~G.}\ \bibnamefont {LeDuc}},
  \bibinfo {author} {\bibfnamefont {B.~A.}\ \bibnamefont {Mazin}}, \bibinfo
  {author} {\bibfnamefont {A.}~\bibnamefont {Vayonakis}}, \ and\ \bibinfo
  {author} {\bibfnamefont {J.}~\bibnamefont {Zmuidzinas}},\ }\href {\doibase
  10.1038/nature02037} {\bibfield  {journal} {\bibinfo  {journal} {Nature}\
  }\textbf {\bibinfo {volume} {425}},\ \bibinfo {pages} {817} (\bibinfo {year}
  {2003})}\BibitemShut {NoStop}%
\bibitem [{\citenamefont {Vora}\ \emph {et~al.}(2012)\citenamefont {Vora},
  \citenamefont {Kumaravadivel}, \citenamefont {Nielsen},\ and\ \citenamefont
  {Du}}]{Vora2011}%
  \BibitemOpen
  \bibfield  {author} {\bibinfo {author} {\bibfnamefont {H.}~\bibnamefont
  {Vora}}, \bibinfo {author} {\bibfnamefont {P.}~\bibnamefont {Kumaravadivel}},
  \bibinfo {author} {\bibfnamefont {B.}~\bibnamefont {Nielsen}}, \ and\
  \bibinfo {author} {\bibfnamefont {X.}~\bibnamefont {Du}},\ }\href {\doibase
  10.1063/1.3703117} {\bibfield  {journal} {\bibinfo  {journal} {Applied
  Physics Letters}\ }\textbf {\bibinfo {volume} {100}},\ \bibinfo {pages}
  {153507} (\bibinfo {year} {2012})}\BibitemShut {NoStop}%
\bibitem [{\citenamefont {Yan}\ \emph {et~al.}(2012)\citenamefont {Yan},
  \citenamefont {Kim}, \citenamefont {Elle}, \citenamefont {Sushkov},
  \citenamefont {Jenkins}, \citenamefont {Milchberg}, \citenamefont {Fuhrer},\
  and\ \citenamefont {Drew}}]{Yan2011}%
  \BibitemOpen
  \bibfield  {author} {\bibinfo {author} {\bibfnamefont {J.}~\bibnamefont
  {Yan}}, \bibinfo {author} {\bibfnamefont {M.-H.}\ \bibnamefont {Kim}},
  \bibinfo {author} {\bibfnamefont {J.~A.}\ \bibnamefont {Elle}}, \bibinfo
  {author} {\bibfnamefont {A.~B.}\ \bibnamefont {Sushkov}}, \bibinfo {author}
  {\bibfnamefont {G.~S.}\ \bibnamefont {Jenkins}}, \bibinfo {author}
  {\bibfnamefont {H.~M.}\ \bibnamefont {Milchberg}}, \bibinfo {author}
  {\bibfnamefont {M.~S.}\ \bibnamefont {Fuhrer}}, \ and\ \bibinfo {author}
  {\bibfnamefont {H.~D.}\ \bibnamefont {Drew}},\ }\href {\doibase
  10.1038/nnano.2012.88} {\bibfield  {journal} {\bibinfo  {journal} {Nature
  nanotechnology}\ }\textbf {\bibinfo {volume} {7}},\ \bibinfo {pages} {472}
  (\bibinfo {year} {2012})}\BibitemShut {NoStop}%
\bibitem [{\citenamefont {Betz}\ \emph {et~al.}(2012)\citenamefont {Betz},
  \citenamefont {Vialla}, \citenamefont {Brunel}, \citenamefont {Voisin},
  \citenamefont {Picher}, \citenamefont {Cavanna}, \citenamefont {Madouri},
  \citenamefont {F\`eve}, \citenamefont {Berroir}, \citenamefont
  {Pla\ifmmode~\mbox{\c{c}}\else \c{c}\fi{}ais},\ and\ \citenamefont
  {Pallecchi}}]{Betz2012}%
  \BibitemOpen
  \bibfield  {author} {\bibinfo {author} {\bibfnamefont {A.~C.}\ \bibnamefont
  {Betz}}, \bibinfo {author} {\bibfnamefont {F.}~\bibnamefont {Vialla}},
  \bibinfo {author} {\bibfnamefont {D.}~\bibnamefont {Brunel}}, \bibinfo
  {author} {\bibfnamefont {C.}~\bibnamefont {Voisin}}, \bibinfo {author}
  {\bibfnamefont {M.}~\bibnamefont {Picher}}, \bibinfo {author} {\bibfnamefont
  {A.}~\bibnamefont {Cavanna}}, \bibinfo {author} {\bibfnamefont
  {A.}~\bibnamefont {Madouri}}, \bibinfo {author} {\bibfnamefont
  {G.}~\bibnamefont {F\`eve}}, \bibinfo {author} {\bibfnamefont {J.-M.}\
  \bibnamefont {Berroir}}, \bibinfo {author} {\bibfnamefont {B.}~\bibnamefont
  {Pla\ifmmode~\mbox{\c{c}}\else \c{c}\fi{}ais}}, \ and\ \bibinfo {author}
  {\bibfnamefont {E.}~\bibnamefont {Pallecchi}},\ }\href {\doibase
  10.1103/PhysRevLett.109.056805} {\bibfield  {journal} {\bibinfo  {journal}
  {Phys. Rev. Lett.}\ }\textbf {\bibinfo {volume} {109}},\ \bibinfo {pages}
  {056805} (\bibinfo {year} {2012})}\BibitemShut {NoStop}%
\bibitem [{\citenamefont {Fong}\ and\ \citenamefont {Schwab}(2012)}]{Fong2012}%
  \BibitemOpen
  \bibfield  {author} {\bibinfo {author} {\bibfnamefont {K.}~\bibnamefont
  {Fong}}\ and\ \bibinfo {author} {\bibfnamefont {K.}~\bibnamefont {Schwab}},\
  }\href {\doibase 10.1103/PhysRevX.2.031006} {\bibfield  {journal} {\bibinfo
  {journal} {Physical Review X}\ }\textbf {\bibinfo {volume} {2}},\ \bibinfo
  {pages} {4} (\bibinfo {year} {2012})}\BibitemShut {NoStop}%
\bibitem [{\citenamefont {Santavicca}\ \emph {et~al.}(2011)\citenamefont
  {Santavicca}, \citenamefont {Chudow}, \citenamefont {Prober}, \citenamefont
  {Purewal},\ and\ \citenamefont {Kim}}]{Santavicca2011}%
  \BibitemOpen
  \bibfield  {author} {\bibinfo {author} {\bibfnamefont {D.~F.}\ \bibnamefont
  {Santavicca}}, \bibinfo {author} {\bibfnamefont {J.~D.}\ \bibnamefont
  {Chudow}}, \bibinfo {author} {\bibfnamefont {D.~E.}\ \bibnamefont {Prober}},
  \bibinfo {author} {\bibfnamefont {M.~S.}\ \bibnamefont {Purewal}}, \ and\
  \bibinfo {author} {\bibfnamefont {P.}~\bibnamefont {Kim}},\ }\href {\doibase
  10.1063/1.3593500} {\bibfield  {journal} {\bibinfo  {journal} {Applied
  Physics Letters}\ }\textbf {\bibinfo {volume} {98}},\ \bibinfo {pages}
  {223503} (\bibinfo {year} {2011})}\BibitemShut {NoStop}%
\bibitem [{\citenamefont {Chudow}\ \emph {et~al.}(2012)\citenamefont {Chudow},
  \citenamefont {Santavicca}, \citenamefont {McKitterick}, \citenamefont
  {Prober},\ and\ \citenamefont {Kim}}]{Chudow2012}%
  \BibitemOpen
  \bibfield  {author} {\bibinfo {author} {\bibfnamefont {J.~D.}\ \bibnamefont
  {Chudow}}, \bibinfo {author} {\bibfnamefont {D.~F.}\ \bibnamefont
  {Santavicca}}, \bibinfo {author} {\bibfnamefont {C.~B.}\ \bibnamefont
  {McKitterick}}, \bibinfo {author} {\bibfnamefont {D.~E.}\ \bibnamefont
  {Prober}}, \ and\ \bibinfo {author} {\bibfnamefont {P.}~\bibnamefont {Kim}},\
  }\href {\doibase 10.1063/1.4704152} {\bibfield  {journal} {\bibinfo
  {journal} {Applied Physics Letters}\ }\textbf {\bibinfo {volume} {100}},\
  \bibinfo {pages} {163503} (\bibinfo {year} {2012})}\BibitemShut {NoStop}%
\bibitem [{\citenamefont {Fu}\ \emph {et~al.}(2008)\citenamefont {Fu},
  \citenamefont {Zannoni}, \citenamefont {Chan}, \citenamefont {Adams},
  \citenamefont {Nicholson}, \citenamefont {Polizzi},\ and\ \citenamefont
  {Yngvesson}}]{Fu2008}%
  \BibitemOpen
  \bibfield  {author} {\bibinfo {author} {\bibfnamefont {K.}~\bibnamefont
  {Fu}}, \bibinfo {author} {\bibfnamefont {R.}~\bibnamefont {Zannoni}},
  \bibinfo {author} {\bibfnamefont {C.}~\bibnamefont {Chan}}, \bibinfo {author}
  {\bibfnamefont {S.~H.}\ \bibnamefont {Adams}}, \bibinfo {author}
  {\bibfnamefont {J.}~\bibnamefont {Nicholson}}, \bibinfo {author}
  {\bibfnamefont {E.}~\bibnamefont {Polizzi}}, \ and\ \bibinfo {author}
  {\bibfnamefont {K.~S.}\ \bibnamefont {Yngvesson}},\ }\href {\doibase
  10.1063/1.2837188} {\bibfield  {journal} {\bibinfo  {journal} {Applied
  Physics Letters}\ }\textbf {\bibinfo {volume} {92}},\ \bibinfo {pages}
  {033105} (\bibinfo {year} {2008})}\BibitemShut {NoStop}%
\bibitem [{\citenamefont {Horng}\ \emph {et~al.}(2011)\citenamefont {Horng},
  \citenamefont {Chen}, \citenamefont {Geng}, \citenamefont {Girit},
  \citenamefont {Zhang}, \citenamefont {Hao}, \citenamefont {Bechtel},
  \citenamefont {Martin}, \citenamefont {Zettl}, \citenamefont {Crommie},
  \citenamefont {Shen},\ and\ \citenamefont {Wang}}]{Horng2011}%
  \BibitemOpen
  \bibfield  {author} {\bibinfo {author} {\bibfnamefont {J.}~\bibnamefont
  {Horng}}, \bibinfo {author} {\bibfnamefont {C.-F.}\ \bibnamefont {Chen}},
  \bibinfo {author} {\bibfnamefont {B.}~\bibnamefont {Geng}}, \bibinfo {author}
  {\bibfnamefont {C.}~\bibnamefont {Girit}}, \bibinfo {author} {\bibfnamefont
  {Y.}~\bibnamefont {Zhang}}, \bibinfo {author} {\bibfnamefont
  {Z.}~\bibnamefont {Hao}}, \bibinfo {author} {\bibfnamefont {H.~A.}\
  \bibnamefont {Bechtel}}, \bibinfo {author} {\bibfnamefont {M.}~\bibnamefont
  {Martin}}, \bibinfo {author} {\bibfnamefont {A.}~\bibnamefont {Zettl}},
  \bibinfo {author} {\bibfnamefont {M.~F.}\ \bibnamefont {Crommie}}, \bibinfo
  {author} {\bibfnamefont {Y.~R.}\ \bibnamefont {Shen}}, \ and\ \bibinfo
  {author} {\bibfnamefont {F.}~\bibnamefont {Wang}},\ }\href {\doibase
  10.1103/PhysRevB.83.165113} {\bibfield  {journal} {\bibinfo  {journal} {Phys.
  Rev. B}\ }\textbf {\bibinfo {volume} {83}},\ \bibinfo {pages} {165113}
  (\bibinfo {year} {2011})}\BibitemShut {NoStop}%
\bibitem [{\citenamefont {Voutilainen}\ \emph {et~al.}(2011)\citenamefont
  {Voutilainen}, \citenamefont {Fay}, \citenamefont {H\"akkinen}, \citenamefont
  {Viljas}, \citenamefont {Heikkil\"a},\ and\ \citenamefont
  {Hakonen}}]{Voutilainen2011}%
  \BibitemOpen
  \bibfield  {author} {\bibinfo {author} {\bibfnamefont {J.}~\bibnamefont
  {Voutilainen}}, \bibinfo {author} {\bibfnamefont {A.}~\bibnamefont {Fay}},
  \bibinfo {author} {\bibfnamefont {P.}~\bibnamefont {H\"akkinen}}, \bibinfo
  {author} {\bibfnamefont {J.~K.}\ \bibnamefont {Viljas}}, \bibinfo {author}
  {\bibfnamefont {T.~T.}\ \bibnamefont {Heikkil\"a}}, \ and\ \bibinfo {author}
  {\bibfnamefont {P.~J.}\ \bibnamefont {Hakonen}},\ }\href {\doibase
  10.1103/PhysRevB.84.045419} {\bibfield  {journal} {\bibinfo  {journal} {Phys.
  Rev. B}\ }\textbf {\bibinfo {volume} {84}},\ \bibinfo {pages} {045419}
  (\bibinfo {year} {2011})}\BibitemShut {NoStop}%
\bibitem [{\citenamefont {Tse}\ and\ \citenamefont
  {Das~Sarma}(2009)}]{Tse2009}%
  \BibitemOpen
  \bibfield  {author} {\bibinfo {author} {\bibfnamefont {W.-K.}\ \bibnamefont
  {Tse}}\ and\ \bibinfo {author} {\bibfnamefont {S.}~\bibnamefont
  {Das~Sarma}},\ }\href {\doibase 10.1103/PhysRevB.79.235406} {\bibfield
  {journal} {\bibinfo  {journal} {Phys. Rev. B}\ }\textbf {\bibinfo {volume}
  {79}},\ \bibinfo {pages} {235406} (\bibinfo {year} {2009})}\BibitemShut
  {NoStop}%
\bibitem [{\citenamefont {Schmidt}, \citenamefont {Schoelkopf},\ and\
  \citenamefont {Cleland}(2004)}]{Schmidt2004}%
  \BibitemOpen
  \bibfield  {author} {\bibinfo {author} {\bibfnamefont {D.}~\bibnamefont
  {Schmidt}}, \bibinfo {author} {\bibfnamefont {R.}~\bibnamefont {Schoelkopf}},
  \ and\ \bibinfo {author} {\bibfnamefont {A.}~\bibnamefont {Cleland}},\ }\href
  {http://prl.aps.org/abstract/PRL/v93/i4/e045901} {\bibfield  {journal}
  {\bibinfo  {journal} {Phys. Rev. Lett.}\ }\textbf {\bibinfo {volume} {93}}
  (\bibinfo {year} {2004})}\BibitemShut {NoStop}%
\bibitem [{\citenamefont {Irwin}(1995)}]{Irwin1995}%
  \BibitemOpen
  \bibfield  {author} {\bibinfo {author} {\bibfnamefont {K.~D.}\ \bibnamefont
  {Irwin}},\ }\href {\doibase 10.1063/1.113674} {\bibfield  {journal} {\bibinfo
   {journal} {Applied Physics Letters}\ }\textbf {\bibinfo {volume} {66}},\
  \bibinfo {pages} {1998} (\bibinfo {year} {1995})}\BibitemShut {NoStop}%
\bibitem [{\citenamefont {Prober}(1993)}]{Prober1993}%
  \BibitemOpen
  \bibfield  {author} {\bibinfo {author} {\bibfnamefont {D.~E.}\ \bibnamefont
  {Prober}},\ }\href {\doibase 10.1063/1.109445} {\bibfield  {journal}
  {\bibinfo  {journal} {Applied Physics Letters}\ }\textbf {\bibinfo {volume}
  {62}},\ \bibinfo {pages} {2119} (\bibinfo {year} {1993})}\BibitemShut
  {NoStop}%
\bibitem [{\citenamefont {Borzenets}\ \emph {et~al.}(2011)\citenamefont
  {Borzenets}, \citenamefont {Coskun}, \citenamefont {Jones},\ and\
  \citenamefont {Finkelstein}}]{Borzenets2011}%
  \BibitemOpen
  \bibfield  {author} {\bibinfo {author} {\bibfnamefont {I.~V.}\ \bibnamefont
  {Borzenets}}, \bibinfo {author} {\bibfnamefont {U.~C.}\ \bibnamefont
  {Coskun}}, \bibinfo {author} {\bibfnamefont {S.~J.}\ \bibnamefont {Jones}}, \
  and\ \bibinfo {author} {\bibfnamefont {G.}~\bibnamefont {Finkelstein}},\
  }\href {\doibase 10.1103/PhysRevLett.107.137005} {\bibfield  {journal}
  {\bibinfo  {journal} {Phys. Rev. Lett.}\ }\textbf {\bibinfo {volume} {107}},\
  \bibinfo {pages} {137005} (\bibinfo {year} {2011})}\BibitemShut {NoStop}%
\bibitem [{\citenamefont {Du}, \citenamefont {Skachko},\ and\ \citenamefont
  {Andrei}(2008)}]{Du2008a}%
  \BibitemOpen
  \bibfield  {author} {\bibinfo {author} {\bibfnamefont {X.}~\bibnamefont
  {Du}}, \bibinfo {author} {\bibfnamefont {I.}~\bibnamefont {Skachko}}, \ and\
  \bibinfo {author} {\bibfnamefont {E.~Y.}\ \bibnamefont {Andrei}},\ }\href
  {\doibase 10.1103/PhysRevB.77.184507} {\bibfield  {journal} {\bibinfo
  {journal} {Phys. Rev. B}\ }\textbf {\bibinfo {volume} {77}},\ \bibinfo
  {pages} {184507} (\bibinfo {year} {2008})}\BibitemShut {NoStop}%
\bibitem [{\citenamefont {Ojeda-Aristizabal}\ \emph {et~al.}(2009)\citenamefont
  {Ojeda-Aristizabal}, \citenamefont {Ferrier}, \citenamefont {Gu\'eron},\ and\
  \citenamefont {Bouchiat}}]{Ojeda-Aristizabal2009}%
  \BibitemOpen
  \bibfield  {author} {\bibinfo {author} {\bibfnamefont {C.}~\bibnamefont
  {Ojeda-Aristizabal}}, \bibinfo {author} {\bibfnamefont {M.}~\bibnamefont
  {Ferrier}}, \bibinfo {author} {\bibfnamefont {S.}~\bibnamefont {Gu\'eron}}, \
  and\ \bibinfo {author} {\bibfnamefont {H.}~\bibnamefont {Bouchiat}},\ }\href
  {\doibase 10.1103/PhysRevB.79.165436} {\bibfield  {journal} {\bibinfo
  {journal} {Phys. Rev. B}\ }\textbf {\bibinfo {volume} {79}},\ \bibinfo
  {pages} {165436} (\bibinfo {year} {2009})}\BibitemShut {NoStop}%
\bibitem [{\citenamefont {Tinkham}(2004)}]{Tinkham2004}%
  \BibitemOpen
  \bibfield  {author} {\bibinfo {author} {\bibfnamefont {M.}~\bibnamefont
  {Tinkham}},\ }\href
  {http://books.google.com/books/about/Introduction\_to\_Superconductivity.html?id=k6AO9nRYbioC}
  {\emph {\bibinfo {title} {{Introduction to Superconductivity}}}},\ \bibinfo
  {edition} {2nd}\ ed.\ (\bibinfo  {publisher} {Dover Publications},\ \bibinfo
  {year} {2004})\ p.\ \bibinfo {pages} {480}\BibitemShut {NoStop}%
\bibitem [{\citenamefont {{Das Sarma}}\ \emph {et~al.}(2011)\citenamefont {{Das
  Sarma}}, \citenamefont {Adam}, \citenamefont {Hwang},\ and\ \citenamefont
  {Rossi}}]{DasSarma2011}%
  \BibitemOpen
  \bibfield  {author} {\bibinfo {author} {\bibfnamefont {S.}~\bibnamefont {{Das
  Sarma}}}, \bibinfo {author} {\bibfnamefont {S.}~\bibnamefont {Adam}},
  \bibinfo {author} {\bibfnamefont {E.}~\bibnamefont {Hwang}}, \ and\ \bibinfo
  {author} {\bibfnamefont {E.}~\bibnamefont {Rossi}},\ }\href {\doibase
  10.1103/RevModPhys.83.407} {\bibfield  {journal} {\bibinfo  {journal}
  {Reviews of Modern Physics}\ }\textbf {\bibinfo {volume} {83}},\ \bibinfo
  {pages} {407} (\bibinfo {year} {2011})}\BibitemShut {NoStop}%
\bibitem [{\citenamefont {Chen}\ and\ \citenamefont {Clerk}(2012)}]{Chen2012}%
  \BibitemOpen
  \bibfield  {author} {\bibinfo {author} {\bibfnamefont {W.}~\bibnamefont
  {Chen}}\ and\ \bibinfo {author} {\bibfnamefont {A.~A.}\ \bibnamefont
  {Clerk}},\ }\href {\doibase 10.1103/PhysRevB.86.125443} {\bibfield  {journal}
  {\bibinfo  {journal} {Phys. Rev. B}\ }\textbf {\bibinfo {volume} {86}},\
  \bibinfo {pages} {125443} (\bibinfo {year} {2012})}\BibitemShut {NoStop}%
\bibitem [{\citenamefont {Mather}(1982)}]{Mather1982}%
  \BibitemOpen
  \bibfield  {author} {\bibinfo {author} {\bibfnamefont {J.~C.}\ \bibnamefont
  {Mather}},\ }\href {\doibase 10.1364/AO.21.001125} {\bibfield  {journal}
  {\bibinfo  {journal} {Applied Optics}\ }\textbf {\bibinfo {volume} {21}},\
  \bibinfo {pages} {1125} (\bibinfo {year} {1982})}\BibitemShut {NoStop}%
\bibitem [{\citenamefont {Dicke}(1946)}]{Dicke1946}%
  \BibitemOpen
  \bibfield  {author} {\bibinfo {author} {\bibfnamefont {R.~H.}\ \bibnamefont
  {Dicke}},\ }\href {\doibase 10.1063/1.1770483} {\bibfield  {journal}
  {\bibinfo  {journal} {Review of Scientific Instruments}\ }\textbf {\bibinfo
  {volume} {17}},\ \bibinfo {pages} {268} (\bibinfo {year} {1946})}\BibitemShut
  {NoStop}%
\bibitem [{\citenamefont {Burke}\ \emph {et~al.}(1996)\citenamefont {Burke},
  \citenamefont {Schoelkopf}, \citenamefont {Prober}, \citenamefont {Skalare},
  \citenamefont {McGrath}, \citenamefont {Bumble},\ and\ \citenamefont
  {LeDuc}}]{Burke1996}%
  \BibitemOpen
  \bibfield  {author} {\bibinfo {author} {\bibfnamefont {P.~J.}\ \bibnamefont
  {Burke}}, \bibinfo {author} {\bibfnamefont {R.~J.}\ \bibnamefont
  {Schoelkopf}}, \bibinfo {author} {\bibfnamefont {D.~E.}\ \bibnamefont
  {Prober}}, \bibinfo {author} {\bibfnamefont {A.}~\bibnamefont {Skalare}},
  \bibinfo {author} {\bibfnamefont {W.~R.}\ \bibnamefont {McGrath}}, \bibinfo
  {author} {\bibfnamefont {B.}~\bibnamefont {Bumble}}, \ and\ \bibinfo {author}
  {\bibfnamefont {H.~G.}\ \bibnamefont {LeDuc}},\ }\href {\doibase
  10.1063/1.116052} {\bibfield  {journal} {\bibinfo  {journal} {Applied Physics
  Letters}\ }\textbf {\bibinfo {volume} {68}},\ \bibinfo {pages} {3344}
  (\bibinfo {year} {1996})}\BibitemShut {NoStop}%
\bibitem [{\citenamefont {Callen}\ and\ \citenamefont
  {Welton}(1951)}]{Callen1951}%
  \BibitemOpen
  \bibfield  {author} {\bibinfo {author} {\bibfnamefont {H.}~\bibnamefont
  {Callen}}\ and\ \bibinfo {author} {\bibfnamefont {T.}~\bibnamefont
  {Welton}},\ }\href {\doibase 10.1103/PhysRev.83.34} {\bibfield  {journal}
  {\bibinfo  {journal} {Physical Review}\ }\textbf {\bibinfo {volume} {83}},\
  \bibinfo {pages} {34} (\bibinfo {year} {1951})}\BibitemShut {NoStop}%
\bibitem [{\citenamefont {Letzter}\ and\ \citenamefont
  {Webster}(1970)}]{Letzter1970}%
  \BibitemOpen
  \bibfield  {author} {\bibinfo {author} {\bibfnamefont {S.}~\bibnamefont
  {Letzter}}\ and\ \bibinfo {author} {\bibfnamefont {N.}~\bibnamefont
  {Webster}},\ }\href {\doibase 10.1109/MSPEC.1970.5213514} {\bibfield
  {journal} {\bibinfo  {journal} {IEEE Spectrum}\ }\textbf {\bibinfo {volume}
  {7}},\ \bibinfo {pages} {67} (\bibinfo {year} {1970})}\BibitemShut {NoStop}%
\bibitem [{\citenamefont {{Ho Eom}}\ \emph {et~al.}(2012)\citenamefont {{Ho
  Eom}}, \citenamefont {Day}, \citenamefont {LeDuc},\ and\ \citenamefont
  {Zmuidzinas}}]{HoEom2012}%
  \BibitemOpen
  \bibfield  {author} {\bibinfo {author} {\bibfnamefont {B.}~\bibnamefont {{Ho
  Eom}}}, \bibinfo {author} {\bibfnamefont {P.~K.}\ \bibnamefont {Day}},
  \bibinfo {author} {\bibfnamefont {H.~G.}\ \bibnamefont {LeDuc}}, \ and\
  \bibinfo {author} {\bibfnamefont {J.}~\bibnamefont {Zmuidzinas}},\ }\href
  {\doibase 10.1038/nphys2356} {\bibfield  {journal} {\bibinfo  {journal}
  {Nature Physics}\ }\textbf {\bibinfo {volume} {8}},\ \bibinfo {pages} {623}
  (\bibinfo {year} {2012})}\BibitemShut {NoStop}%
\bibitem [{\citenamefont {DeFeo}\ and\ \citenamefont
  {Plourde}(2012)}]{DeFeo2012}%
  \BibitemOpen
  \bibfield  {author} {\bibinfo {author} {\bibfnamefont {M.~P.}\ \bibnamefont
  {DeFeo}}\ and\ \bibinfo {author} {\bibfnamefont {B.~L.~T.}\ \bibnamefont
  {Plourde}},\ }\href {\doibase 10.1063/1.4742164} {\bibfield  {journal}
  {\bibinfo  {journal} {Applied Physics Letters}\ }\textbf {\bibinfo {volume}
  {101}},\ \bibinfo {pages} {052603} (\bibinfo {year} {2012})}\BibitemShut
  {NoStop}%
\bibitem [{\citenamefont {Kittel}(2004)}]{Kittel2004}%
  \BibitemOpen
  \bibfield  {author} {\bibinfo {author} {\bibfnamefont {C.}~\bibnamefont
  {Kittel}},\ }\href {http://books.google.com/books?id=kym4QgAACAAJ} {\emph
  {\bibinfo {title} {{Introduction to Solid State Physics}}}}\ (\bibinfo
  {publisher} {Wiley},\ \bibinfo {year} {2004})\ p.\ \bibinfo {pages}
  {704}\BibitemShut {NoStop}%
\end{thebibliography}
\end{document}